\documentclass[preprint2]{aastex}
\newcommand{\ie}{{\rm i.e. }}
\newcommand{\rmi}{{\rm i}}
\newcommand{\rmj}{{\rm j}}

\shorttitle{Lunar orbital neutron data}
\shortauthors{Eke et al.}
\begin{document}
\title{A quantitative comparison of lunar orbital neutron data}
\author{V.R. Eke}
\affil{Institute for Computational Cosmology, Department of Physics,
  Durham University, South Road, Durham. DH1 3LE, UK}
\email{v.r.eke@durham.ac.uk}

\author{L.F.A. Teodoro}
\affil{BAER, Planetary Systems Branch, Space Sciences and Astrobiology Division, MS 245-3, NASA Ames Research Center, Moffett Field, CA 94035, USA}

\author{D.J. Lawrence}
\affil{Johns Hopkins University Applied Physics Laboratory, Laurel, MD 20723, USA}

\author{R.C. Elphic}
\affil{NASA Ames Research Center, Moffett Field, CA 94035, USA}

\author{W.C. Feldman}
\affil{Planetary Science Institute, 1700 East Fort Lowell, Suite 106, Tucson, AZ 85719, USA}

\begin{abstract}
Data from the Lunar Exploration Neutron Detector (LEND) Collimated
Sensors for Epithermal Neutrons (CSETN) are used in conjunction with a model
based on results from the
Lunar Prospector mission to quantify the extent of the background in
the LEND CSETN. A simple likelihood analysis implies that at least $90\%$
of the lunar component of the LEND CSETN flux results from high energy
epithermal neutrons passing through the walls of the collimator. Thus,
the effective full-width at half-maximum of the LEND CSETN field of view
is comparable with that of the omni-directional Lunar Prospector Neutron
Spectrometer. The resulting map of high energy epithermal neutrons
offers the opportunity to probe the hydrogen abundance at low latitudes, and provide 
constraints on the distribution of lunar water.
\end{abstract}

\keywords{Moon --- methods: data analysis}

\section{Introduction}

The weathering of the lunar surface is so slow that it should provide
extractable information about 
the history of volatile molecule transport within the inner Solar System, a
subject of considerable relevance to water-based life on Earth.
Samples of lunar regolith returned in the Apollo missions
contained a sufficiently low abundance of water that these molecules
were, until recently, dismissed as terrestrial contamination
\citep{eps73} and the Moon was considered to be dry. The change of
paradigm has been driven by evidence that has accumulated over
the past 15 years from a number of different missions and types of
experiment \citep{law11b}. 

The first of this
recent wave of techniques for divining for lunar water was
the use of bistatic circularly polarised radar measurements from the
Clementine mission \citep{noz96}. A peak at
opposition in the ratio of returned same-sense to opposite sense
circularly polarised flux was 
posited to be due to multiple scattering within a low-loss
medium such as water ice. The signal correlated with Clementine being
over the lunar poles, the locations of permanently shaded `cold traps'
in which water ice would be stable against sublimation for billions of
years. \cite{sta97} and \cite{cam06} found that the same signal
would also result from a surface that was rough on the scale of the
$13$cm wavelength of the radiation and that even in sunlit
regions this same high circular polarisation ratio was seen. Thus it
is unclear whether or not water ice deposits are the explanation for
these observations. Theoretical modelling of the radar scattering
process by \cite{fa11} suggests that this technique will struggle
to detect ice mixed in with lunar regolith because of their similar
relative permittivities. However, \cite{spu10} argue, based upon
measurements from Chandrayaan-1's Mini-SAR experiment,
that surface roughness cannot explain the signal seen from within some
`anomalous' polar craters where no signal is seen from the surrounding
region, which might be expected to have had post-cratering roughness
weathered away at a similar rate.

An important method that has provided evidence relevant to the
discussion of lunar water over the last 13 years is neutron spectroscopy.
It was realised 50 years ago that the neutron flux leaking from the
Moon encoded information about the hydrogen content in the lunar
surface \citep{lin61}. Free neutrons are created via
interactions between incoming cosmic rays and nuclei within the top
few metres of the lunar regolith. Having been knocked out of nuclei,
these neutrons subsequently evolve by scattering off other
nuclei until they either escape through the surface or have
sufficiently little kinetic energy that they are absorbed by another
nucleus. The fast neutrons (energies exceeding $0.5$~MeV)
predominantly lose energy through inelastic collisions that
result in gamma rays being emitted from the temporarily excited nuclei
that they hit. At intermediate energies ($0.3$~eV$<E<0.5$~MeV),
epithermal neutrons donate kinetic energy to their targets, which
recoil following 
elastic collisions. As hydrogen nuclei have a similar mass to that of
the neutron, they are particularly effective at draining energy
rapidly from epithermal neutrons and sending them down to thermal
energies ($E<0.3$~eV), where they become increasingly susceptible
to nuclear capture. Thus, a hydrogen rich top metre of regolith would
be expected to produce a deficit in the epithermal neutron flux
leaking from the lunar surface.

This first calculation of the anticipated sensitivity of orbital
neutron measurements to the regolith hydrogen content was refined by
\cite{metz90} and \cite{bill91} with the benefit of improved knowledge
of the regolith composition. These models were confronted
with observational evidence when results from Lunar Prospector's
Neutron Spectrometer were produced \citep{bill98b}. A 
convincing deficit of epithermal neutrons was detected in the
vicinity of both lunar poles. However, Lunar Prospector (LP) was an
omni-directional detector that accepted neutrons from all
directions. Consequently, the spatial resolution of the
published hydrogen maps was sufficiently poor that the question of how
well these count rate dips correlated with surface 
features, such as craters, remained tantalisingly unanswered.
Other results from LP's Neutron and Gamma Ray Spectrometers
\citep{david98,bill98a,bill98b,rick98} provided a detailed and coherent
picture of the lunar surface composition. The analysis of the full 
data set \citep{bill00,bill01}, along with a detailed account of the data
reduction process \citep{sly04} and improved modelling of both the
detector response and variation with composition \citep{law06} provide
a benchmark against which current and future experiments can be compared.

A recent development in the story of lunar
water has come courtesy of measurements of the
infra-red reflectance of sunlit regions of the Moon. Three
different studies of the infra-red spectrum \citep{cla09,sun09,pie09}
found evidence for absorption by oxygen
to hydrogen chemical bonds on the lunar surface. Furthermore, the
depth of the absorption features appeared to vary with solar zenith
angle. Such a result would imply the existence of a contribution by
hydroxyl and water molecules to the dynamic, if tenuous, lunar
exosphere. This idea of a lunar water cycle had been considered by
\cite{but97} and \cite{cv00}.

The most compelling evidence for the existence of lunar water came
from the Lunar Crater Observation and Sensing Satellite (LCROSS),
which trailed a 
spent Centaur rocket into a permanently shaded part of the Cabeus
crater near to the lunar south pole. Nine instruments on LCROSS
analysed the plume of material resulting from this impact. It was
determined to have contained water molecules at $5.6\pm2.9$ per cent
by mass \citep{cola10}. This uncertainty is dominated by the
uncertainty in the total plume mass, and the detection of water vapour
and water ice within the instrumental field of view was a much more
significant $155\pm12$kg.

Given that water does exist on the Moon, the question of how it came
to be there becomes relevant. \cite{arn79} considered various
potential sources: delivery by comets or meteorites, solar wind
reduction of iron in the lunar regolith or degassing from the lunar interior.
As a result of the slow weathering of the lunar surface, the quantity,
isotopic ratios and distribution of water, both with depth and 
position on the surface, should still contain information about the
history of water delivery to the Moon. It is thus important to
make an accurate map of the distribution of lunar water in order to help
disentangle the importance of the various delivery options.

With this aim in mind, there are two different approaches that have been
taken to try to improve upon the original LP hydrogen distribution
maps. The software route involves using image reconstruction techniques
to undo the instrumental blurring by the omni-directional LP detector,
whereas the hardware route is to make the detector itself collimated.

\cite{rick07} introduced
a pixon-based image reconstruction algorithm that extracted more information
from the LP epithermal neutron data set, and \cite{me09}
showed that, when all permanently shaded regions at both poles were
stacked together, these data statistically favoured a reconstruction
where the hydrogen was concentrated into the permanently shaded
regions. This result has important implications for the molecular form
of the hydrogen, its method of delivery to the Moon and survival on
the surface. With improved maps of the permanently shaded polar regions
from the Japanese SELENE mission \citep{noda08}, \cite{luis10}
confirmed this earlier result and made lunar polar hydrogen maps that
were given to the LCROSS team to help them decide where to crash their
Centaur rocket in the search for water ice \citep{cola10,schultz}.

The alternative approach, to deploy hardware that only accepts neutrons from a
small fraction of $4\pi$ steradians, was taken with the Lunar
Exploration Neutron Detector (LEND), described by \cite{mit08}, on the
Lunar Reconnaissance Orbiter (LRO). Blocking epithermal neutrons from other
directions is a challenge best met by placing many moderating and
absorbing nuclei in the way. However, the necessary requirement not to
send large masses away from the 
Earth opposes this straightforward approach to producing a collimated
neutron detector. An attempt to balance these competing constraints
has been made with the 4 LEND Collimated Sensors for Epithermal Neutrons
(CSETN1-CSETN4), hereafter referred to collectively as CSETN.
This is one of seven instruments \citep{chin07} on board the LRO mission
that has, to date, been mapping the Moon since mid-2009.

There has been some disagreement in the literature concerning the
sensitivity of the LEND CSETN. \cite{law10} argue that the
detected flux should contain about 0.18 collimated neutrons per
second. This conclusion was arrived at following both analytical
calculations, using 
geometrical considerations to scale the LP count rate,
and Monte Carlo neutron transport modelling, which also provided an
estimate of the uncollimated lunar background flux of 0.09 counts per
second. These findings were based upon the not entirely specific
description of the detector configuration provided by
\cite{mit08}. The estimate for the collimated neutron count rate
given by \cite{mit08} is 0.9 counts per second. They also
state that the total background from all sources should provide about
1.2 neutrons per second of background counts. More recently, using data 
gathered by the LEND CSETN during the first year in orbit around the Moon,
\cite{mit10b} assert that the collimated neutron count rate is more
like 1.9 neutrons per second, with 0.3 counts per second of
uncollimated lunar signal and 2.8 neutrons per second as a result of
cosmic ray strikes on LRO itself. However, \cite{law11a} use the
compositional variation of the high- and low-energy epithermal neutron
count rates to estimate that the uncollimated lunar flux exceeds 2
counts per second, leaving very little collimated signal.

The abundance and distribution of near-surface
lunar hydrogen provides a valuable constraint on models describing the
delivery and survival of volatile materials over billions of
years. Making accurate maps of the lunar hydrogen distribution
is thus one of the foremost aims for those seeking to understand this
important aspect of the history of the inner solar system.
As the ability to detect small dips in epithermal 
neutron flux is paramount when using orbital neutron data to infer
hydrogen abundances in the top $\sim 1$m of lunar regolith, it is
crucial to know precisely what is the signal-to-background ratio 
for the LEND CSETN. The most pertinent
question to address is what fraction
of the detected neutrons actually originate from the lunar surface in
the small field of view of the collimator? The remainder of the
detected neutrons would be uncollimated and originating either in the
Moon or in the spacecraft itself.

The Geosciences Node of NASA's Planetary Data System 
(PDS\footnote{http://pds-geosciences.wustl.edu}) 
is a public repository that archives and distributes digital data
related to the study of the Moon, and it contains data from the 
LEND CSETN. It is thus timely to compare
these new data with those collected over a decade ago by the LP
Neutron Spectrometer. The count rate
variation with spacecraft position and altitude provides a direct
means to decompose the total detected flux into collimated signal and
uncollimated background noise.
The main purpose of this paper is to answer the question, what
fraction of the LEND CSETN counts are epithermal neutrons that truly
originate from the small piece of lunar surface geometrically
accessible to the collimator?


Section~\ref{sec:method} of this paper provides a description of 
the data that 
will be used in this study and the likelihood analysis that is
employed to determine what fraction of the LEND CSETN count rate is
collimated epithermal neutrons. Section~\ref{sec:res} contains the
results of using the data to decompose the total counts into those
from the various components, as well as a discussion of how robust
these findings are to potential shortcomings in the modelling of the
count rate. These results are 
compared with the component fractions advocated by \cite{mit10b}.
The findings are summarised in Section~\ref{sec:conc}.

\section{Method}\label{sec:method}

\begin{figure*}
\hspace{-0.4cm}\includegraphics[width=17.cm]{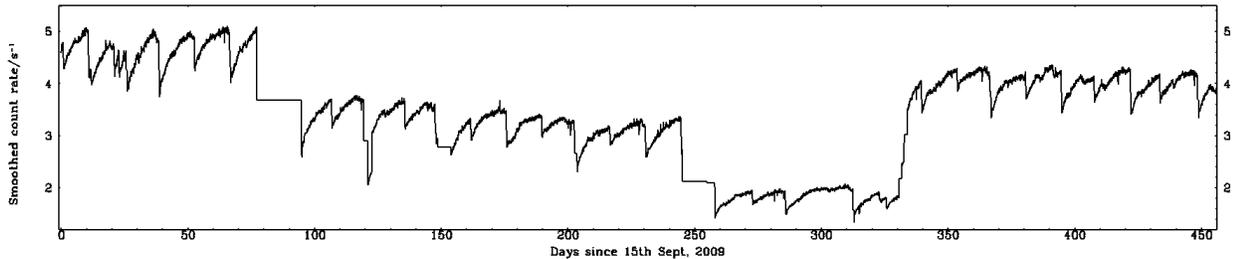}
\vspace{-0cm}
\caption{The box-car smoothed (with a box containing 10000 
consecutive 1 second observations) LEND CSETN data for the first 15 months 
of operation.
\label{fig:ts}}
\end{figure*}

\begin{figure}
\includegraphics[width=8.cm]{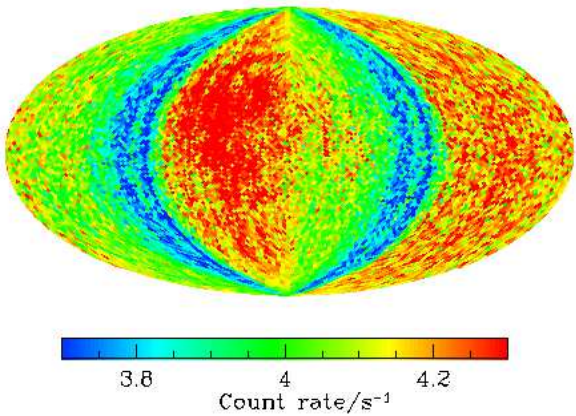}
\caption{A Mollweide projection of the raw LEND CSETN channel 10-16 data.
  This uses the HEALPix pixellation \citep{healp} at a resolution of 
$N_{\rm side}=64$, which corresponds to a pixel side length of 
$\sim 28$ km. 
\label{fig:glitch}}
\end{figure}

This section contains a detailed description of the data being
used, the model that is employed to describe them, and the procedure
used to find the best-fitting model parameters. These parameters are
the fraction of the total LEND CSETN counts
coming from each of the different possible sources, namely collimated
lunar neutrons, uncollimated lunar neutrons that have enough energy to
pass through the sides of the collimator and uncollimated
spacecraft neutrons.

\subsection{Data}

The data analysed in this paper are the first 15 months of `rectified
science data' available within the LEND Reduced Data Record (RDR) data
product available at the Geosciences Node of NASA's PDS.
Data from all 4 LEND CSETN detectors
are studied. The neutron count rate comprises the sum of counts in
channels 10-16, because these channels mostly contain neutron detections rather
than the gamma rays or charged particles that prevail in channels 1-9
(Mitrofanov, private communication). 
To avoid needing to know detailed information
concerning neutron transport within LRO and LEND, only the observations
for which the nadir angle is less than $5^\circ$ are used. In this
configuration, the LEND instrument lies slightly beneath and to one
side of the LRO spacecraft and has an unobstructed view of the
lunar surface. Other than on the first day of collection of these data
(15th September, 2009), LRO was in a reasonably circular mapping orbit
with altitudes generally between 40 km and 60 km. During the initial
day of data, the spacecraft had yet to be put into its mapping orbit,
so reached altitudes as low as 30 km and as high as 210 km. Throughout
the rest of this paper, the zero of time will be defined as the start
of the data collection period.

\subsubsection{Long-term temporal variations in the time series data}

The box-car smoothed time series data are shown in Fig.~\ref{fig:ts}. It is
apparent that there are some glitches in these smoothed count
rates. These are almost always associated with periods when no data
are available from the detectors, and have a periodicity of roughly 2
weeks, or approximately half a lunar month. After the glitches, the count rate
can be as much as $20\%$ lower than before, and it gradually rises
over a period as long as a few weeks. Fig.\ref{fig:glitch} shows a map
of the raw count rate. It is apparent that there are
bands of low counts at $-110^\circ<$longitude$<-90^\circ$ and
$70^\circ<$longitude$<90^\circ$, and bands of high count rate at
$-80^\circ<$longitude$<-0^\circ$ and
$100^\circ<$longitude$<180^\circ$. The detectors are orbiting in a
plane whose normal is approximately the Earth-Moon axis when they are switched off.

These artefacts are largely removed when
the temporal variability is modelled as described in the next section.
Recorded count rates are sometimes unusually high both slightly
before and after the glitch times, so data are removed in the
periods 5 minutes prior to the time when the data stop and 1 minute
after the time when they reappear. 

The one example of a glitch where the count rate jumps upwards is
found at $t\sim 122.5$ days, within a short period between two other
glitches. To avoid any problems this might introduce, all data
with $[120<t/$days $<123]$ are discarded.

Other notable features of Fig.~\ref{fig:ts} are that there are gaps
without data at $t\sim 80$ and $250$ days, after which the flux is
significantly lower. As described on the PDS, this is because CSETN3
and CSETN2 respectively are switched off at these times. Both of these
detectors recommence recording data around $t\sim 335$ days. Along
with the temporal variation described above, this should also be
accounted for when trying to predict the total counts for any
particular observation.

The number of 1s observations remaining after pruning the data as
described above is $\sim 35$ million. They span a total of 456 days from
15th September, 2009. Various subsets of these
data are used in Section~\ref{sec:res} to investigate the robustness
of the results, but the default in this paper is to use all of these
observations.

\subsection{Model}\label{ssec:mod}

There are 3 components that will contribute to the LEND CSETN channel
10-16 count rate. The magnitudes of each component will depend upon a
combination of the spacecraft latitude, longitude and altitude,
$a$. The components are:
\begin{flushleft}
1) collimated lunar epithermal neutrons,\\
2) uncollimated lunar neutrons, with energies high enough to make it
through the walls of the collimator,\\
3) spacecraft-derived neutrons also entering into the detectors.\\
\end{flushleft}
The final component may also include any contribution to the channel 10-16 
count rate due to charged particles resulting from cosmic rays striking the 
spacecraft.

The $^3$He tubes used to count incoming neutrons in LEND, and also
LP, are covered in cadmium to keep out low energy, thermal 
neutrons at energies below $\sim 0.4$eV. The energy dependence of both
the detector efficiency \citep{law02} and the 
lunar neutron leakage flux \citep{mck06} means that low energy
epithermal (LEE) neutrons with $\sim 0.4 - 100$eV provide the bulk of
the counts in these detectors. In addition to the Cd coating, the LEND
CSETN are placed 
inside a collimator that consists of layers of polyethylene to
moderate the neutrons and $^{10}$B to absorb the moderated
neutrons. \cite{law11a} estimate that the collimator is 
largely transparent for neutrons that strike it with at least $\sim
10$keV. Incident neutrons at these higher energies could be moderated
to a low enough energy that the $^3$He counter is
able to detect them. It is this route that neutrons comprising
components (2) and (3) must take in order to contribute to the LEND
CSETN count rate. Component (2) reflects the high energy
epithermal (HEE) neutron flux from the lunar surface; an interesting
science product in itself.

The technique employed here essentially constructs a model for the
total count rate that the LEND CSETN should measure in each 1s
observation from each of the components. This model is based upon
the LP maps of fast, moderated, epithermal and thermal neutron flux
and a spatially invariant spacecraft 
background. Each of the model component fluxes also requires the
inclusion of both an altitude dependence and the temporal count rate
gain variation,
$g(t)$. Once the individual components have their latitude, longitude,
altitude and time variations computed, they are each 
normalised such that a fraction $f_\rmi$ of the total counts summed
over all observations in the data set comes from component \rmi. In
this nomenclature, determining $f_\rmi$
is the purpose of this paper. The LEND CSETN data are used to
determine the relative probabilities of the different sets of
component fractions, $f_\rmi$, \ie they determine which component
fractions provide the best match to the time series measurements. 

The evaluation of the spatial, altitude and temporal variations of
the model component fluxes are described in the following subsections.

\subsubsection{Constructing the spatial variation in the model}\label{sssec:spa}

\begin{figure}
\includegraphics[width=8.cm]{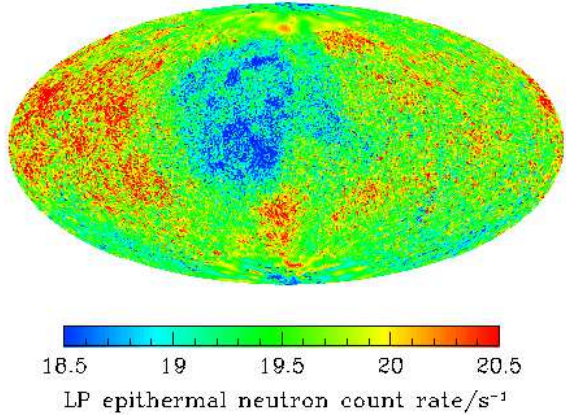}
\caption{
A Mollweide projection of the LP epithermal neutron
map including the pixon reconstructions in the polar regions, which is
used as the template for component (1) of the model. This uses the
HEALPix pixellation at a resolution of $N_{\rm side}=64$, which
corresponds to a pixel side length of $\sim 28$ km. 
}
\label{fig:lpepi}
\end{figure}

\begin{figure}
\includegraphics[width=8.cm]{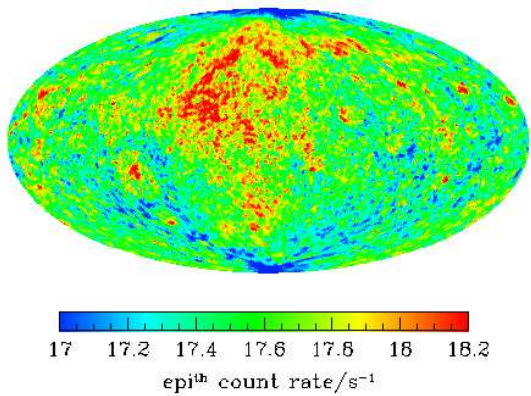}
\caption{
A Mollweide projection of the thorium-corrected epi$^{tho}$
map, used as part of component (2) in the model. This uses the HEALPix
pixellation at a resolution of $N_{\rm side}=64$.
}
\label{fig:thorium}
\end{figure}

\begin{figure}
\includegraphics[width=8.cm]{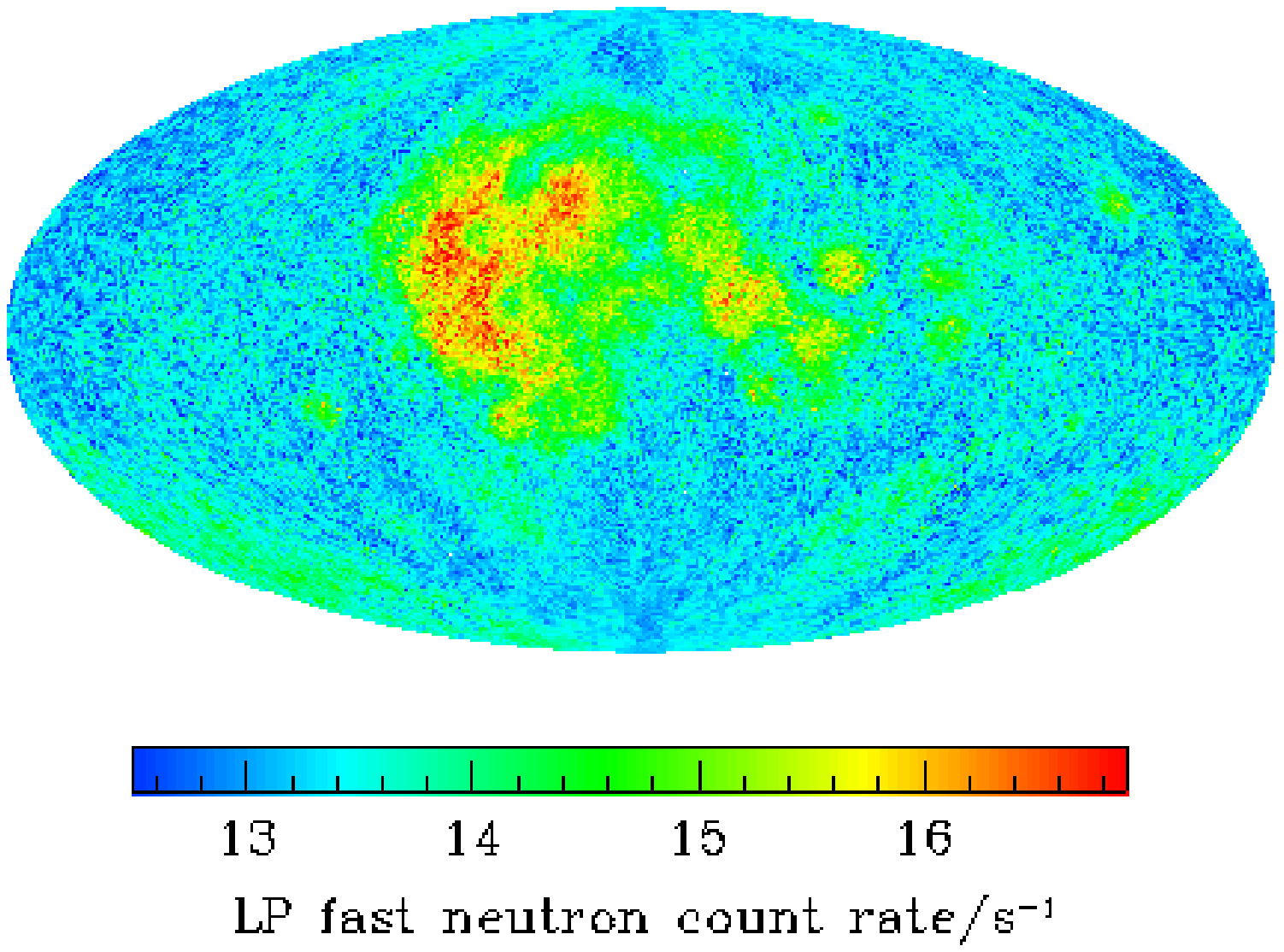}
\caption{
A Mollweide projection of the LP fast neutron
map, used as part of component (2) in the model. This uses the HEALPix
pixellation at a resolution of $N_{\rm side}=64$.
}
\label{fig:lpfast}
\end{figure}

Given the well documented and well understood LP maps for the spatial
variation of the thermal, epithermal and fast neutron fluxes, these data
sets are the natural resource for designing whole-Moon templates of the 
various components of the LEND CSETN flux. Each component of the model
requires a template that describes how the relative count rate varies with
longitude and latitude. These templates will hereafter be denoted with
$u_\rmi$ for component $\rmi$, and their normalisation is discussed at
the end of subsection~\ref{ssec:mod}.
To model the spatial variation of the components of
lunar origin, various different resolution maps of the LP
neutron data are made using the HEALPix
pixellation scheme \citep{healp}. This partitioning of the sphere
has 12 equal area base pixels and finer pixellations are achieved by
splitting each of the pixels into 4 equal area pixels. The level of
resolution can be denoted by $N_{\rm side}$, the number of pixels
along a side of a base pixel. In this nomenclature, the total number
of pixels describing the spherical surface is $N_{\rm pix}=12 N_{\rm
  side}^2$ and, for the Moon, the typical pixel side length for the
kth level of refinement is $l_{\rm k}=(1778/N_{\rm side})$ km.

Given that LRO can be at altitudes from $30-210$ km, although it
spends most of the time at its nominal mapping altitude of $50$ km,
it is necessary to choose the appropriate resolution template, when
calculating the model counts for a particular observation. This is
done quite crudely by linearly interpolating between the pixel values
either side of the FWHM for the component under consideration. There
are inevitable inaccuracies in this treatment, both because LP did not
have comparable spatial resolution to component (1), and it did not
produce a HEE map containing the same spectrum of neutron energies as
would be expected to produce component (2). However, this simple approach
will still provide an adequate comparison between LP and LEND CSETN
data sets. 

It is necessary to find an expression for the
variation of the FWHM with altitude, so that an appropriate
pixellation can be chosen to give a suitably smoothed measure of the
lunar flux. This can be achieved by considering
the neutron flux emanating from a small area on the
lunar surface. If the detector is orbiting at an altitude $a$, a
distance $r$ from the neutron source under consideration, with a
corresponding arc length $s$ between the sub-spacecraft point and the
source, then the relative flux received can be written as
\begin{equation}
\frac{f(s)}{f(s=0)}=\left(\frac{a}{r}\right)^2 (\cos\theta)^{1+\alpha},
\end{equation}
where $\alpha$ represents the effective beaming of neutrons relative to the
constant surface brightness, Lambertian case, and $\theta$ is the
angle from the surface normal to the spacecraft. This `beaming' arises because 
the epithermal neutron number density increases with depth in the top layer 
of the regolith \citep{mck06}, and is analogous to solar limb darkening.
Using Monte Carlo
neutron transport models, \cite{law06} found that $\alpha\sim 0.5$
provided a good description of the beaming of epithermal neutrons from
the lunar surface. Simple trigonometry yields the following expression for
the change in flux with sub-spacecraft distance along the surface,
from which the altitude dependence of the FWHM can be calculated:
\begin{equation}
\frac{f(s)}{f(s=0)}=\left(\frac{a}{r_{\rm M}}\right)^2 \frac{[y
    \cos(s/r_{\rm M})-1]^{1+\alpha}}{[y^2+1-2y \cos(s/r_{\rm M})]^\frac{3+\alpha}{2}},
\label{psf}
\end{equation}
where $r_{\rm M}=1737.4$ km represents the lunar radius and $y=(r_{\rm
M}+a)/r_{\rm M}$. This expression assumes that the detector has no
preferred direction. While not strictly true, comparison of the
inferred LP point-spread function (PSF) with that of \cite{sly04} suggests 
that it is an adequate description in that case.

Having described how the component templates will be used, the next
step involves actually creating them.
The template for the collimated LEE component should look like a
noiseless, high spatial resolution version of the LP epithermal neutron map.
While the spatial resolution of the LP epithermal map is that of an
omni-directional detector, in the polar regions, the pixon
reconstructions of the LP epithermal data \citep{luis10} can be used
to gain higher spatial resolution for the LEE component (1) template,
$u_1$, as shown in Fig.~\ref{fig:lpepi}. Using the pixon maps allows
the model to be extended from a resolution of $N_{\rm side}=64$ to
$N_{\rm side}=128$ and $256$ at $|{\rm lat}|>71^\circ$. This highest
resolution roughly corresponds to the anticipated footprint for
component (1) in the model. The results do not depend greatly on this
detail in the model, as demonstrated in Section~\ref{sec:res}.

The spatial variation of the HEE component (2) is somewhat less easy to
estimate, both because the relevant incident neutron energies depend
upon precise details of the collimator, which are not available, and
there does not exist a high-fidelity map of lunar neutron emission
integrated with the appropriate energy-dependent filter.
However, neutron data from the LP Gamma-ray
Spectrometer (GRS) anticoincidence shield \citep{gen03} were used to
derive information from epithermal neutrons in the energy range
$100-500$ keV, which has significant overlap with the LEND HEE
neutrons. These `moderated' neutrons showed a strong correlation with
LP-measured fast neutrons \citep{sly00}, which themselves are
correlated with average atomic mass \citep{gas01}. The measured
correlation of moderated neutrons with fast neutrons suggests that the
HEE neutrons may also, in part, be correlated with fast neutrons. The
LP moderated neutron data are consistent with neutron transport models
that show HEE neutrons are highly correlated with fast neutrons and
average atomic mass \citep{law11a} in regions with low concentrations
of hydrogen. For regions with enhanced hydrogen concentrations,
modelled HEE neutrons \citep{law11a} and measured moderated neutrons
\citep{gen03} show a decrease in neutron count rates that is very
similar to LEE neutrons.

Based on this information, a variety of educated guesses can be made
for a HEE template by
considering all of the LP thermal, epithermal, moderated and fast
neutron 
maps. In this study, two different types of initial HEE template have
been devised. 
For the first template, a map similar to the epi$^*$
constructed by \cite{bill98b} is created by subtracting $6.4\%$ of 
the LP thermal count rate from the LP epithermal map, which
effectively removes the LEE compositional variation due to neutron
absorption at energies below $100$ eV. To account for a fast neutron
contribution as described above, a
to-be-determined amount of the LP fast neutron map is added. The
precise fraction of fast neutrons depends on details of the LEND
construction, and is therefore left for the data to select.

An option of correcting for residual, nearside neutron absorption in
the epi$^*$ map is also considered by subtracting a portion of the LP
thorium concentrations \citep{law03} that spatially correlate with the
residual neutron absorptions.
Fig.~\ref{fig:thorium} shows the template for this
epi$^{tho}$ component of the HEE variation and Fig.~\ref{fig:lpfast}
shows the LP fast neutron map, which also contributes to $u_2$, the
template for component (2).

The second method
involves replacing all regions of the LP epithermal map where the LP
thermal counts are lower than $20$ neutrons per second with a scaled
version of the LP fast neutron map. This is essentially targetting the
mare regions, where the LEE and thermal neutrons have a dip in count
rate, whereas the fast neutrons have an excess as would be expected
from LP moderated neutron measurements and HEE neutron transport
simulations. 
Unlike for component (1), the HEE lunar template only needs to be
constructed at the low resolution corresponding to an uncollimated
detector.

By using only nadir facing observations, the spacecraft background should
be independent of location over the surface, although not independent
of altitude, so $u_3$ is taken to be a constant, independent of
latitude and longitude.

\subsubsection{Constructing the altitude dependence in the
  model}\label{sssec:alt}

\begin{figure}
\includegraphics[width=7.5cm]{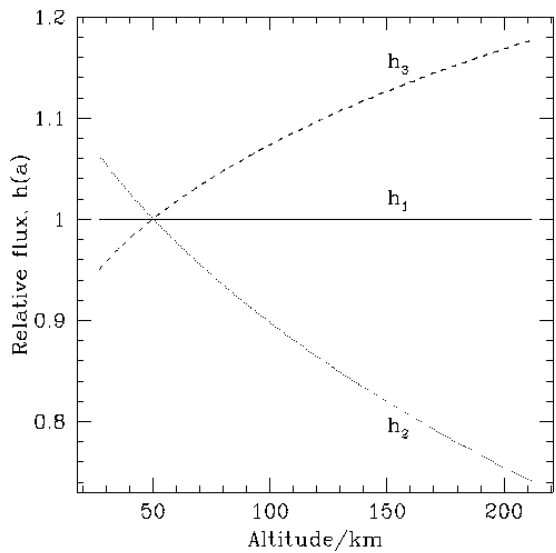}
\caption{
The altitude dependence, $h(a)$, of the count rates for the various different
components. A solid line traces the variation with altitude of the
collimated LEE lunar flux. Uncollimated HEE lunar flux, $h_2$, drops
off as shown by the dotted line. The spacecraft flux, $h_3$, is traced
by the dashed line. 
}
\label{fig:altdep}
\end{figure}

The altitude dependence of the lunar flux components, $h_1$ and $h_2$,
can be found by integrating eq.~(\ref{psf}) over the appropriate
fraction of the lunar surface out to the horizon for the chosen
altitude, $a$. For the collimated epithermal component, the limit of
the integration surface is approximated as $s=a\beta_{\rm col}$, where
$\beta_{\rm col}=5.6^\circ$ is the half-opening angle of the
collimator. There is almost no altitude variation of the collimated
component because the flux from each visible patch of surface drops
like $a^{-2}$, while the area of surface grows like $a^2$. As the
beaming of neutrons is weak, the Moon presents a disc of almost
uniform neutron surface brightness, these two factors almost entirely
cancel out, provided that the Moon always fills the collimator's field
of view.

The uncollimated lunar HEE component is computed in a
similar way by integrating out to the horizon. In this case, as the
spacecraft altitude increases, the Moon subtends a smaller angle and
the uncollimated flux decreases. One might expect that the beaming of
HEE would differ from that of LEE, because HEE neutrons are more likely to
have been created by cosmic rays hitting the lunar surface at more
grazing angles. The model is given the freedom to choose a value of
$\alpha\ne 0.5$ for component (2), if the data prefer.

The altitude dependence of the spacecraft flux arises as a result of
the solid angle of cosmic ray-containing sky blocked out by the
Moon. Consequently, higher altitudes lead to more
cosmic rays hitting the spacecraft and more locally generated neutrons
being detected by the LEND CSETN. The flux of cosmic rays hitting the
spacecraft as a whole varies like
\begin{equation}
{\rm CRF}\propto \left(1+\sqrt{1-\frac{1}{y^2}}\right).
\label{cralt}
\end{equation}
One might wonder if the flux of cosmic rays hitting the nadir
facing side of LRO is more relevant for predicting the neutron flux
reaching LEND. This varies with altitude more rapidly:
\begin{equation}
{\rm CRF}\propto \sqrt{1-\frac{1}{y^2}}.
\label{underside}
\end{equation}
An experiment was performed where two components of uniform background were
introduced, one with no altitude dependence and another varying like
eq.~(\ref{underside}), these being the extreme limits of how component
(3) might depend upon altitude. The data picked out a most likely
model with a combination of these two components that corresponded to
an overall spacecraft background altitude dependence as given by
eq.~(\ref{cralt}). Thus, the altitude dependence of component (3) was
set to
\begin{equation}
{\rm h_3}\propto \left(1+\sqrt{1-\frac{1}{y^2}}\right).
\end{equation}
Fig.\ref{fig:altdep} shows the altitude dependence of all three model
components of the count rate, with each normalised to unity at $a=50$
km.

\subsubsection{Constructing the temporal variation in the model}\label{sssec:temp}

The variation of the effective count rate gain of the detector must be 
accounted for in the model to describe the received flux as a function of
longitude, latitude, altitude and time. While the reason for the
glitches and gradual recoveries in the count rate is unclear, physical systems
might be expected to have exponentially time-varying behaviour. Thus,
a natural and effective function to fit to each interglitch period is
\begin{equation}
C=C_0+A(1-{\rm e}^{-\lambda(t-t_0)}),
\label{glitch}
\end{equation}
where $C$ represents the count rate, $t$, the time, $C_0$ is the count
rate at a reference time $t_0$ in this interglitch period, and $A$ and
$\lambda$ are constants to be determined by fitting to the data. This
approximate correction is applied for the first evaluation of the most
likely component fractions. It takes into account the exponential
recoveries in count rate and the sharp drops across periods when the
detectors are not returning measurements. If there are any cosmic ray
flux variations on timescales of a week or longer, then they will also be
absorbed into this empirical temporal correction.

Any temporally correlated residual differences, on timescales shorter
than about a week, between the most likely model and the data are assumed to
result from cosmic ray flux variations. These are estimated using the
boxcar smoothed data and model over periods of 40ks, and this `cosmic
ray' factor can also be placed into the temporal variation in
the model. It is, of course, possible that inexact glitch removal can
lead to incorrectly inferred cosmic ray variation, as these two
factors are essentially degenerate. Only the assumption that the
glitches are described by eq.~(\ref{glitch}) decouples the two.
\cite{mit10b} do not mention the glitches in the data
despite the fact that they require systematic corrections more than an order 
of magnitude greater than the quoted statistical uncertainties in their 
results.
A comparison of the inferred relative cosmic ray flux with an independent 
proxy, namely the terrestrial neutron flux,
is given in Section~\ref{sec:res}.

To take into account the combination of the
long-term variations in the data, count rates for each model component
are normalised to produce the observed total number of counts over the
entire data set when $f_\rmi=1$. This corresponds to 3.58
counts/second. As a result, each observation time has its own
normalisation factor, defined such that the measured count rate is a
factor $g(t)$ times what it would have been without these temporal
variations associated with the detectors. For the initial likelihood
evaluation, $g(t)$ only corrects for the varying number of detectors
and the glitches in the data, whereas for the second likelihood
evaluation it also includes the cosmic ray correction factor.

Taking into account all of these different components of the modelled
count rate, the anticipated count rate for any given observation, j,
can be written as 
\begin{equation}
m_\rmj=g(t_\rmj)\sum_{\rmi=1}^{3} f_{\rmi}h_\rmi(a_\rmj)u_\rmi(t_\rmj).
\end{equation}
The free parameters are the component fractions, $f_\rmi$, determining the
relative contributions of the components. In fact only 2 of these are
independent because the sum of all fractions is $1$ in order to
recover the correct mean neutron count rate.

\subsection{Likelihood evaluation}\label{ssec:like}

By choosing a particular set of fractions, $\{f_\rmi\}$, it is possible
to determine what is the probability of measuring the time series LEND
CSETN data set. Varying $\{f_\rmi\}$ allows the most likely set of
component fractions to be inferred.
The probability of measuring $d_\rmj$ counts in observation j, where the model
predicts $m_\rmj$ counts, is given by the Poisson distribution:
\begin{equation}
P_\rmj=\frac{m_\rmj^{d_\rmj}e^{-m_\rmj}}{d_\rmj!}.
\end{equation}
If the likelihood of measuring a set of $n_{\rm obs}$ data values is
denoted by 
\begin{equation}
L=\prod_{\rmj=1}^{n_{\rm obs}} P_\rmj,
\end{equation}
then, ignoring the model-independent denominator, a log-likelihood can
be defined as
\begin{equation}
\ln{L}=\sum_{\rmj=1}^{n_{\rm obs}}(d_\rmj\ln(m_\rmj)-m_\rmj).
\end{equation}
The log-likelihood is calculated throughout the relevant region of the
component fraction hyperspace, and the most likely parameter values, as defined
by the data, are located. These log-likelihoods can be converted back
to probabilities that the LEND CSETN data would arise given this
model, and normalised so that the probability integrated over
parameter space equals 1. The one-dimensional marginalised probability
can be calculated by integrating these probabilities at fixed values
of any particular parameter, $f_\rmi$. An additional constraint is
that none of the components is allowed to produce a negative
contribution to the count rate.

\section{Results}\label{sec:res}

This section contains the results of applying the method described in
Section~\ref{sec:method} to the LEND CSETN data. Before describing the
inferred component fractions, two additional improvements to the
previously outlined model are described. These are essentially
motivated by information in the data themselves.

\subsection{Additional empirical corrections}

\begin{figure*}
\includegraphics[width=16.5cm]{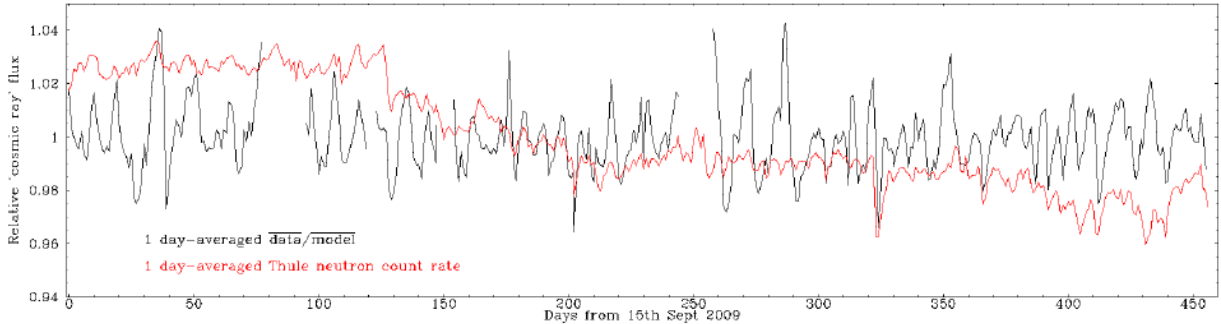}
\caption{Time series showing the variability in the 1-day averaged
  neutron flux at the Thule station of the Bartol Research Institute
  and the ratio of the 40ks box-car smoothed LEND CSETN data to the
  similarly smoothed model. Both of these quantities can be viewed as
  proxies for the variability in the cosmic ray flux.
}
\label{fig:crts}
\end{figure*}

Using the HEE template based upon the combination of the
thermal-corrected LP epithermal and LP fast neutron maps, the entire
15 months of glitch-corrected LEND CSETN were used to determine the
maximum likelihood component fractions. The temporal correlations in
the residuals between the data and most likely model, on time scales
of 40ks, are assumed to arise as a result of variations in the cosmic
ray flux. This implies that, with the residual for each time series
measurement defined as
\begin{equation}
R_\rmj=d_\rmj-m_\rmj,
\end{equation}
periods when the residuals are systematically positive or negative
correspond to times when the cosmic ray flux is high or low
respectively. 

A test of this assumption is shown in Fig.~\ref{fig:crts} with the 
correlation of the ratio of the smoothed LEND CSETN data to the
smoothed model values and the neutron flux measured on Earth at the
Thule station operated by the Bartol Research Institute.
It is apparent in the time series in
Fig.~\ref{fig:crts} that high neutron fluxes are often coincident with
periods where the LEND CSETN data are systematically higher than
the most likely model predicts. Considering the ratio of smoothed data
to smoothed model as
a proxy for cosmic ray flux, the relative mean values for each day can
be compared with the ground-based data. A straightforward test gives a
Spearman rank correlation coefficient of $0.19$, which is highly
significant.

Fig.~\ref{fig:crts} also shows a long-term trend
in the Thule data that is not present in the LEND CSETN cosmic ray
proxy. This is because that temporal variation has been absorbed into
the glitch correction described earlier. Consequently, a fairer
comparison results from normalising the ground-based count rates by their
one-month box-car smoothed values. This effectively removes the long
timescale trends from the Thule data. In this case, the comparison
between the LEND CSETN cosmic ray proxy and the detrended Thule
neutron count rate gives a Spearman coefficient of 0.38. A comparable
test for data from other monitoring stations gives Spearman
coefficients of 0.32 (Swarthmore/Newark), 0.29 (McMurdo) and 0.27
(South Pole), all of which are highly statistically significant.

This evidence provides an important consistency check for the model used here -
namely that one can infer something like the cosmic ray flux
from just the LEND CSETN count rates and the approximate
glitch corrections represented by eq.~(\ref{glitch}).
Using the temporal correlations in the ratio of smoothed data to
smoothed model on scales
of 40ks as a proxy for the variability of the cosmic ray flux,
the correction factor, $g(t_\rmj)$, is multiplied by this cosmic ray
correction factor to remove these correlated residuals,
and the maximum likelihood component fractions are recalculated. 

\begin{figure}
\includegraphics[width=8.cm]{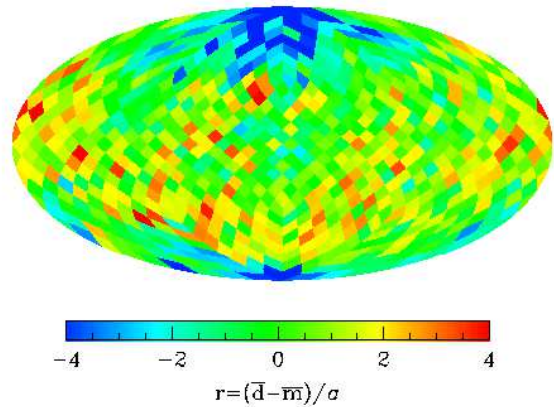}
\caption{The spatial correlation of the reduced residuals (residuals
  normalised by an uncertainty in the pixel, $\sigma$, derived from
  Poisson counting statistics) for the default HEE template, having
  only corrected for the cosmic ray flux variations. Blue regions
  represent areas where the model overestimates the LEND CSETN
  data. $N_{\rm side}=8$ is used for this map, corresponding to pixel
  scales of $\sim 220$ km.
}
\label{fig:resthorl3}
\end{figure}

\begin{figure}
\includegraphics[width=7.5cm]{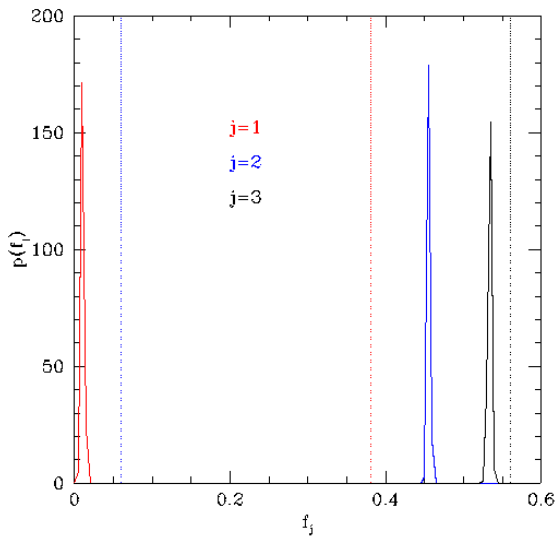}
\caption{
Marginalised probability distributions for the fractions of the total
LEND CSETN counts in components: 1 - collimated lunar LEE neutrons
(red), 2 - uncollimated lunar HEE neutrons (blue), and 3 - locally
produced spacecraft neutrons (black). The coloured vertical dotted
lines are the corresponding values asserted by \cite{mit10b}.
}
\label{fig:marg}
\end{figure}

This second iteration makes very little difference to
the inferred component fractions, although it does significantly
improve the match of the most likely model to the time series
data. A map of the residuals from this second iteration is shown in
Fig.~\ref{fig:resthorl3}. The pixel size is $\sim 220$ km, corresponding to
$N_{\rm side}=8$ in the HEALPix notation. 
While the temporal correlations in the residuals have been removed,
there are still significant spatially correlated residuals. The
collimated flux should have a spatial dependence that is well
described by component (1) in the model, but, as noted in
Section~\ref{sec:method}, the HEE component is more of a challenge.
The most likely reason for these residuals is an
inappropriate choice of model HEE template, so it is updated
to remove all of the residuals measured on the large scales shown
in Fig.~\ref{fig:resthorl3}. The largest correction
factor in the template count rate arising from this change is only
$\sim 3\%$. While this decreases the number of
degrees of freedom provided by the data, it does not preclude the
residuals on smaller scales subsequently being used to test how well
the model describes the data.

In contrast to the temporal correction, which could be cross-checked
against an independent estimate of the cosmic ray flux, there is no
other available map of the Moon with the appropriate energy-dependent
filter applied, whatever that might be in detail.
One reason to think that this is not greatly biasing the results of
this study is that, when starting from the alternative HEE
template described in the previous section, the
residuals are significantly improved relative to those in
Fig.~\ref{fig:resthorl3} and despite this, the inferred component
fractions hardly change. Thus the results found here are robust
to choosing very different input HEE templates.

\begin{figure}
\includegraphics[width=7.5cm]{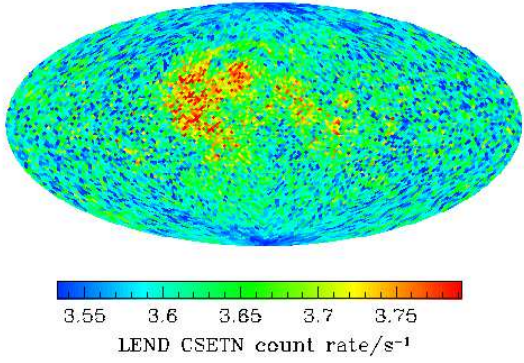}
\caption{
Mollweide projection of the LEND CSETN corrected count rate map,
using $N_{\rm side}=32$, which corresponds to a pixel side length of
$\sim 56$ km.
}
\label{fig:csetn}
\end{figure}

\begin{figure}
\includegraphics[width=7.5cm]{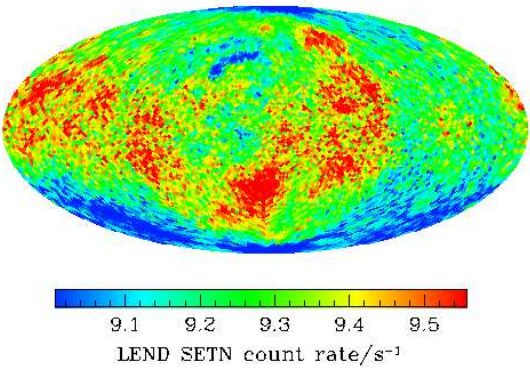}
\caption{
Mollweide projection of the LEND SETN corrected count rate map,
using $N_{\rm side}=32$.
}
\label{fig:setn}
\end{figure}

\subsection{Most likely component fractions}

\begin{figure}
\includegraphics[width=7.5cm]{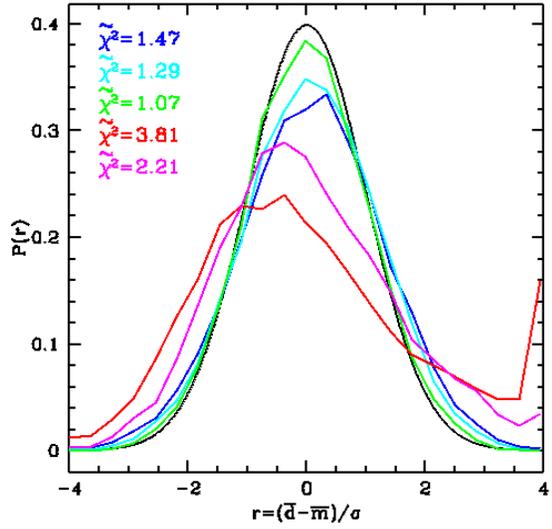}
\caption{Histograms of the reduced residuals, $r$, in pixels at a
  resolution of $N_{\rm side}=32$. The black curve is a Gaussian with
  width 1. Blue, cyan and green curves show the results for the most
  likely model before any corrections, after the cosmic ray
  correction, and after both cosmic ray and HEE template corrections
  respectively. The reduced $\chi^2$ values, accounting for the decrease in
  the number of degrees of freedom caused by the data-based
  corrections, are given in corresponding colours in the
  legend. Results in red and magenta are for the component fractions favoured by
  \cite{mit10b} and \cite{mit11} respectively. 
}
\label{fig:reshist}
\end{figure}

\begin{figure}
\includegraphics[width=7.5cm]{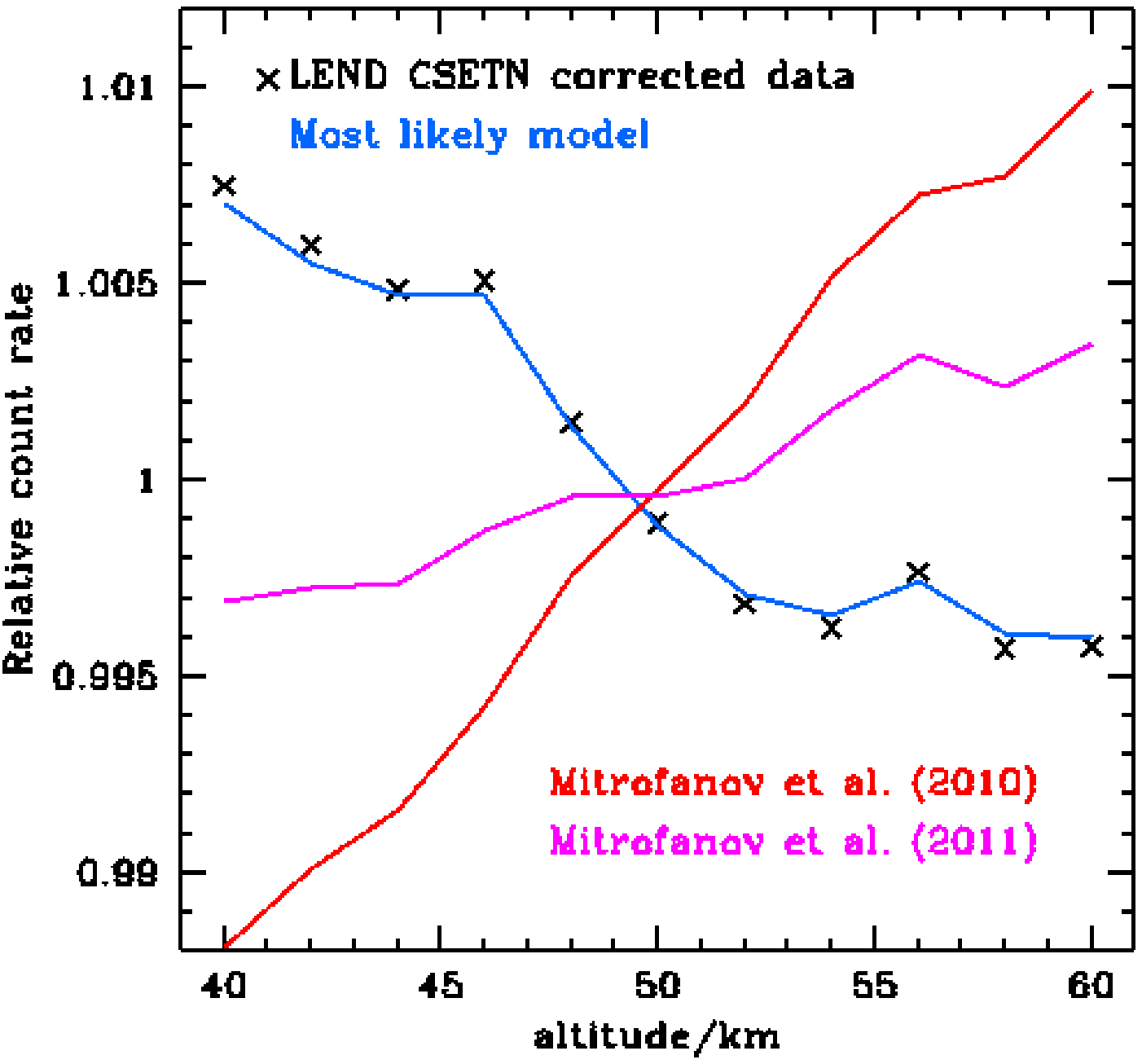}
\caption{The relative count rate as a function of altitude. Points
show the corrected LEND CSETN data (uncertainties are smaller than the
symbol size). The blue, red and magenta lines show the trends from the most
likely model found here and the asserted component fractions of
\cite{mit10b} and \cite{mit11} respectively.
}
\label{fig:altvar}
\end{figure}

Fig.~\ref{fig:marg} shows the marginalised 1D probability
distributions for the component fractions after both the cosmic ray
correction and the HEE map modification. The most likely component
fractions only change from $\{f_1,f_2,f_3\}=\{0.02,0.44,0.54\}$ to
$\{0.01,0.445,0.535\}$ as a result of both of these changes. Statistical
uncertainties on these values, as represented by the widths of the
probability distributions are barely more than $\sim 0.01$. There are
larger systematic uncertainties that are considered in more detail
later in this section. The preferred component fractions reported by
\cite{mit10b} are also shown in this figure. Including the range of
values they quote, their $f_3$ is consistent with that found
here, namely just over half of the LEND CSETN count rate comes from
neutrons originating in the spacecraft itself. However, the division
of the lunar flux between components (1) and (2) is effectively
reversed, with \cite{mit10b} claiming that the collimated flux forms
the vast majority, whereas the LEND CSETN data, coupled with the
simple model based on previous knowledge from LP presented in this
paper, suggests that the lunar flux has roughly ten times as much
coming through the collimator walls in a dominant component (2) as it
does in collimated neutrons.

One might wonder if the reason for this discrepancy is that the LEND
CSETN detector itself perhaps has a very different behaviour to the LP
epithermal neutron detector, despite the fact that they are both
$^3$He-filled Cd-covered tubes. If this were the case, then the model
for component (1) could be in error, leading to incorrect component
fractions. However, if one considers the map of
lunar neutrons produced by the LEND SETN, which is an equivalent
detector strapped to the outside of the LEND collimator, this
explanation can be rejected. Figures~\ref{fig:csetn}
and~\ref{fig:setn} show the corrected (for glitches and cosmic ray
variations) maps for the LEND CSETN and SETN respectively. It is clear
from this comparison that the CSETN map is not a sharp version of that
from the SETN, which is what would be produced if the lunar
variability in the CSETN map were dominated by collimated
neutrons. The lunar dependence of the CSETN map is driven by something
different to the SETN results. This is not consistent with the
asserted component fractions of \cite{mit10b}. Instead it reflects the
preponderance of higher energy neutrons in the CSETN lunar flux. In fact,
\cite{mit11} state that a significant fraction of their `local background from the 
LRO spacecraft' is actually scattered lunar HEE, which will have both a different
spatial and altitude dependence to background arising from cosmic rays striking LRO.
They give $\{f_1,f_2,f_3\}=\{0.32,0.22,0.46\}$ as the component fractions.

One can study the distribution of pixel reduced residuals, $r$, (\ie
residuals normalised by the Poisson counting error in each pixel) on
scales smaller 
than that used for updating the HEE model template to see how well the
most likely models fit the LEND data. Fig.~\ref{fig:reshist} shows
histograms of the pixel residuals for $N_{\rm side}=32$ for the most
likely models before either cosmic ray or HEE template corrections,
after each of these model updates, and for both the \cite{mit10b} and \cite{mit11}
component fractions fed into this corrected model. The most
likely model improves significantly as a result of the temporal and
spatial corrections. However, even before either of them, it was still
vastly superior to either the component fractions of \cite{mit10b} or \cite{mit11}, 
for which the jump in the distribution at $r=4$ contains all pixels with
residuals at least this large. These are essentially pixels in the
mare regions where the data show a high count rate while the
collimated flux should show a decrease relative to other equatorial
locations. The number of degrees of freedom is such that, for a good
fit, the reduced $\chi^2$ should be $1\pm0.01$. Thus, even after both
corrections, the most likely model found here is still not quite a
`good' fit. However, given the approximate treatment of the spatial
component of the modelling on scales of the instrumental PSF, this is
an impressive agreement considering the small number of model
parameters being fitted to the data.

Given the significant difference between the most likely component
fractions found here, and those advocated by \cite{mit10b} and \cite{mit11}, one might
expect that there would be other tests that would allow the data to
choose between these two models. In particular, given that the
uncollimated lunar flux decreases strongly with increasing detector
altitude, whereas the collimated flux is approximately constant, the
altitude dependence of the data should also be a powerful
discriminant. There are correlations of typical spacecraft altitude
with position over the surface, such that LRO is usually lower over
the mare regions than elsewhere, so one needs to account for these by
comparing each time series observation with the model count rate for
that observation. These results, normalised by the mean overall count
rate and binned with respect to altitude, are shown in
Fig.~\ref{fig:altvar}. The data show a slight decrease of count rate
with increasing altitude that can only be reproduced if the model
contains enough uncollimated lunar flux to overcome the opposite trend
brought from the dominant spacecraft background (see
Fig.~\ref{fig:altdep}). Once again, the asserted component fractions
of both \cite{mit10b} and \cite{mit11} are completely inconsistent with the data. 
They imply an altitude dependence that is the opposite of that seen in the
data. This points to the lunar flux predominantly being uncollimated
rather than collimated.

\subsection{Investigation of potential systematic errors}

\begin{table*}
\begin{center}
\caption{Most likely parameter values calculated on a grid with
  spacing $0.005$. Statistical uncertainties are
  of order $0.01$ in all cases, which is smaller
  than the systematic variations arising from different subsets of the
  data, or model assumptions. The default model is for the
  `thorium-corrected' input map. Lines 1-8 represent changes to the
  model, whereas lines 9-21 are for the default model with different
  subsets of the data.}
\begin{tabular}{ccccc}
\tableline
Row number & Experiment & $f_1$ & $f_2$ & $f_3$\\
\tableline
\\
1 & default & $0.010$ &$0.455$&$0.535$ \\
2 & no HEE correction & $0.020$ &$0.450$&$0.530$ \\
3 & no CR correction too & $0.020$ &$0.440$&$0.540$ \\
4 & no thorium & $0.005$ &$0.460$&$0.535$ \\
5 & epi/fast HEE template & $0.010$ &$0.465$&$0.525$ \\
6 & coupled reconstructions & $0.010$ &$0.455$&$0.535$ \\
7 & no pixon reconstructions & $0.005$ &$0.455$&$0.540$ \\
8 & max $N_{\rm side}=128$ & $0.010$ &$0.455$&$0.535$ \\
9 & $0<t/$days$<90$ & $0.000$ & $0.450$ & $0.550$ \\
10 & $90<t/$days$<250$ &$0.040$&$0.435$&$0.525$ \\
11 & $250<t/$days$<335$ &$0.040$&$0.450$&$0.510$ \\
12 & $335<t/$days$<456$ &$0.035$&$0.455$&$0.510$ \\
13 & $a/$km$<45$ &$0.005$ &$0.445$ &$0.550$ \\
14 & $45<a/$km$<55$ &$0.020$ &$0.450$ &$0.530$ \\
15 & $55<a/$km &$0.015$ &$0.445$ &$0.540$ \\
16 & near side &$0.010$ &$0.460$ &$0.530$ \\
17 & far side &$0.005$ &$0.455$ &$0.540$ \\
18 & $|$latitude$|>60^\circ$ & $0.020$ &$0.430$&$0.550$ \\
19 & $|$latitude$|<60^\circ$ & $0.015$ &$0.440$&$0.545$ \\
20 & latitude$>0$ &$0.040$ &$0.415$ &$0.545$ \\
21 & latitude$<0$ &$0.010$ &$0.475$ &$0.515$ \\
\tableline
\end{tabular}
\end{center}
\label{tab:res}
\end{table*}

Given that the statistical uncertainties are so small, it is pertinent
to ask how sensitive are these results to different assumptions
underpinning the model. Table~1 contains results for
various different model assumptions and subsets of the data, from
which it is possible to gain some insight into the robustness of the
inferred component fractions.

The effect of not correcting the HEE template using the residuals is
shown in row 2, whereas neglecting the `cosmic ray' temporal
correction too leads to the fractions in row 3. While these
corrections greatly improve the residuals, as shown in
Fig.~\ref{fig:reshist}, they have no significant impact on the most
likely component fractions. Rows 4 and 5 show the inferred most likely
component fractions for the cases where the initial HEE template does
not include the thorium correction and for the alternative LP epi/fast
template respectively. Once again, the component fractions are robust
to these changes to the model. The same story is repeated for rows 6
and 7, where the decoupled pixon reconstructions of the LP epithermal
map reported in \cite{luis10} are replaced with the coupled ones
produced by \cite{me09}, or just the straightforward binned LP
data. The final aspect of the model that was varied was the minimum
pixel size used to bin the LP data to make the templates for the HEE
model. Rather than using the default $N_{\rm side}=64$, one tier higher
HEALPix resolution was adopted to model the HEE component. Once again,
this made little 
difference, despite gaining some resolution at the cost of introducing
more noise in the model template for the HEE component. Changing the
pixel size for the LEE component also had little effect, because it is
such a sub-dominant component.

These results all provide evidence to support that none of the more
uncertain aspects of the model are greatly impacting upon the
component fractions inferred here. Keeping the default model choices,
the data themselves were then split into different subsets, to see if
similar results were obtained.

Splitting the time series data into the 4 distinct periods within
which the number of active collimators does not change yields the
results in rows 9-12. There seems to be a slight preference for data
taken after $t=90$ days to have a higher fraction in the collimated
component, mostly at the expense of the spacecraft background. It is
difficult to ascertain what aspect of the experiment altered to give
rise to this change, which is statistically significant.

Restricting the data to include only observations made from restricted
ranges in altitude does not appear to make any strong systematic
changes to the inferred component fractions (rows 13-15). Using only
the data for either the near or far side of the Moon yields almost
identical most likely component fractions (rows 16 and 17), and the
results for polar and equatorial regions in rows 18 and 19 are very
similar too.

A significant difference is seen between the component fractions
inferred from northern and southern hemisphere data (rows 20-21), with
the north preferring more collimated and less uncollimated lunar flux
than the south. This presumably reflects residual inaccuracies in the
HEE template. It is reassuring though that, while statistically
significant, these systematic changes in the most likely component
fractions are of a magnitude that is very small relative to the
difference between the values inferred here and those of \cite{mit10b}.

\subsection{Comparison with previous studies}

Breaking down the average $\sim 3.6$ counts per second (over the whole period
used) to count rates in the various components, the results presented
here suggest that on average, these comprise $\sim 0.1$ per second in
collimated epithermals, $\sim 1.6$ in uncollimated lunar HEE, and
$\sim 1.9$ per second in spacecraft
generated counts. Taking into account the variable number of sensors
active during the period, the collimated epithermal count rate is
similar to the result of \cite{law10}. This is not surprising, because
it is based on a simple geometrical argument that scales the flux from
the omni-directional LP experiment to the solid angle subtended by the
collimator.

In the supporting online material for their paper,
\cite{mit10b} perform a similar calculation where they scale the count
rate from the LEND SETN to estimate the collimated lunar flux reaching
the CSETN. However, they neglect the contribution of the
locally generated background to the SETN count rate and overestimate the
beaming of lunar epithermal neutrons. Both of these factors will lead to an
erroneously high anticipated collimated flux.

The neutron beaming can be assessed
using the type of argument that led to eq.~(\ref{psf}). In order to
recover the beaming factor of `about 2' asserted by \cite{mit10b}, the
neutrons would need to be beamed with $\alpha\sim 1.5$, in contrast to
the $\alpha=0.5$ measured from Monte Carlo neutron transport
modelling of the lunar regolith \citep{law06}. This factor of `about 2'
appears to be $50\%$ larger than one would infer for $\alpha=0.5$.

Without an accurate description of the LEND instrument and detailed
modelling of neutron transport within the spacecraft, it is difficult 
to quantify the extent of
the locally-generated background. However, it should be noted that the
SETN is placed next to a large piece of polyethylene,
which is known to moderate and scatter neutrons very effectively
\citep{rick08}. This is the reason it is used as the outer part of the
collimator. Thus it is implausible that there are not significant
numbers of neutrons entering the SETN that come from this
polyethylene, rendering any attempt to scale the SETN count rate to
predict the CSETN count rate of little worth. Comparing
the LEND SETN and LP epithermal maps in Figures~\ref{fig:setn}
and~\ref{fig:lpepi} respectively, it is apparent that they look
qualitatively different in a way that suggests the SETN has an extra
HEE component that decreases the dip in the mare regions.

While the collimated flux found here is in agreement with that
anticipated by \cite{law10}, there are approximately 20 times as many
background lunar HEE neutrons than they predicted based on an
incomplete knowledge of the detector geometry. This
appears to be a result of \cite{law10} having been too optimistic in their
assumption as to how effective the collimator was. Specifically,
\cite{law10} assumed a $^{10}$B density of $2.5$g/cm$^3$, while the
true density was slightly less than $1$g/cm$^3$ \citep{mit10a}. In addition, 
\cite{law10} assumed that except for the opening aperture, the
collimating material completely enclosed the $^3$He sensors.  In contrast, the
true LEND collimator has "holes" that allow neutrons to enter the $^3$He sensor
through the back side of the collimator \citep{mit10a}.
Both of these effects will significantly
increase the true background compared to that estimated by \cite{law10}.
Consequently, one
should view their conclusions concerning the LEND CSETN signal-to-noise
ratios as having been too favourable by a factor of $\sqrt{20}\sim4$. 

\section{Conclusions}\label{sec:conc}

\begin{figure}
\includegraphics[width=7.5cm]{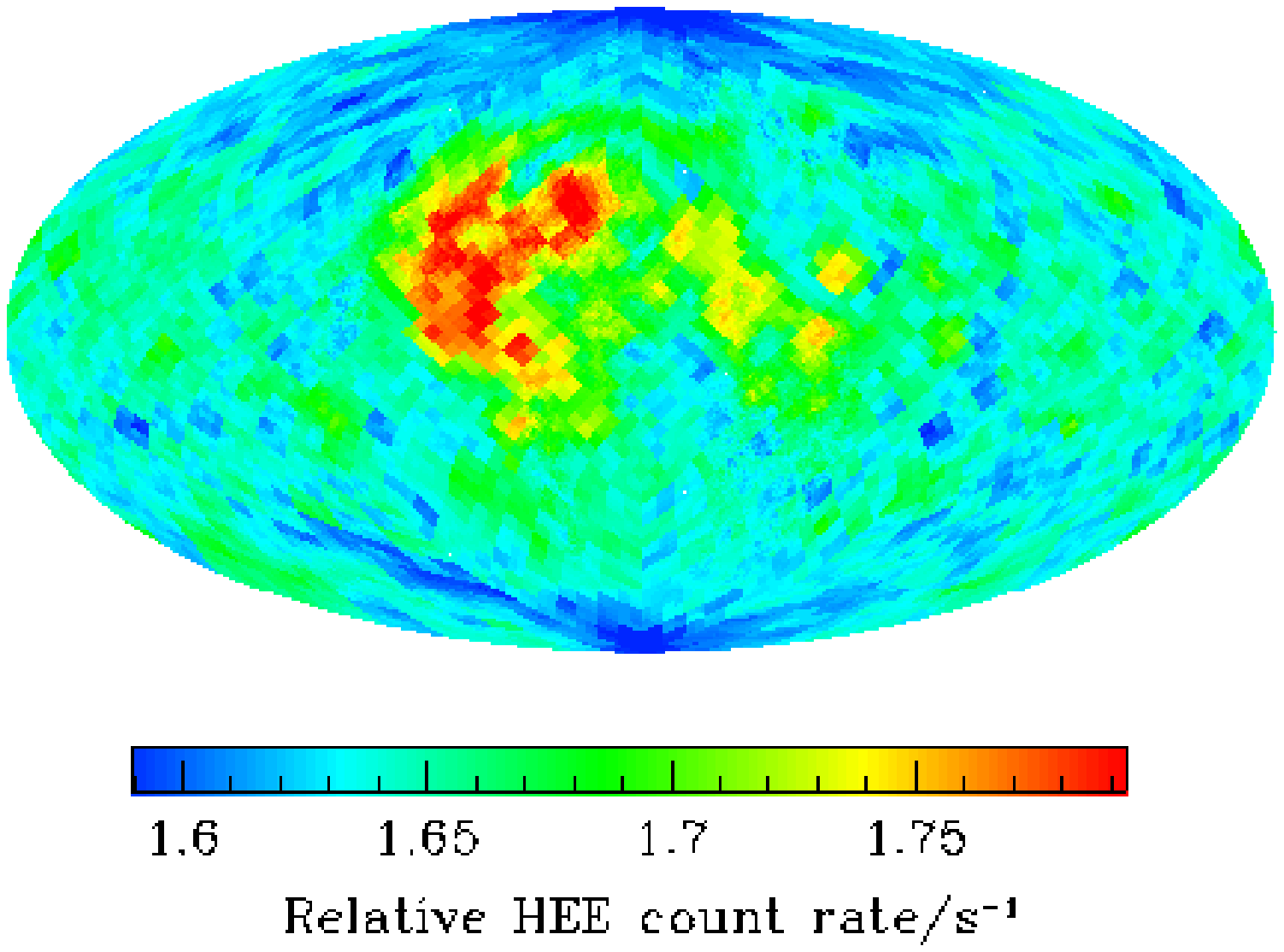}
\caption{
Mollweide projection of the final relative HEE count rate map from the
LEND SHEEN data set using $N_{\rm side}=64$. The arbitrary normalisation is
chosen to match the mean count rate from the HEE component over the
first 456 days of lunar mapping. Note that
the correction for the correlated residuals on scales of $\sim 200$ km
is visible.
}
\label{fig:hee}
\end{figure}

A comprehensive maximum likelihood analysis has been performed on the
LEND CSETN data to determine the sources of the neutrons that it
detects. The fraction of detected neutrons that are collimated lunar
epithermal neutrons is under $0.05$, with the most likely fraction
being $\sim 0.01$. $0.4-0.45$ are uncollimated lunar
HEE neutrons, and the remaining $0.5-0.55$ are neutrons generated in
LRO by direct cosmic ray strikes. The two lunar component fractions
are almost the opposite of those asserted by \cite{mit10b}, who favour
the lunar flux being predominantly collimated. However, the data
themselves are completely inconsistent with this interpretation, both
using the full likelihood analysis and also from a simple inspection
of the global SETN and CSETN maps. The conclusions drawn by
\cite{mit10b} concerning hydrogen distributions should thus be viewed
with skepticism. 

Given that the LEND CSETN is effectively producing an uncollimated map
of the lunar HEE neutrons, albeit with a small contamination from a
LEE collimated component and a large, but spatially uniform spacecraft
background, the experiment would more usefully be called the LEND
Sensor for High Energy Epithermal Neutrons (SHEEN). One would expect
that the HEE neutrons will be less beamed than the LEE ones, because
they are typically created by more grazing-incidence cosmic
rays. This, in combination with the 50 km orbit of LRO relative to 30
km for LP, counteracts the effect of the small fraction of collimated
LEE neutrons, meaning that the effective FWHM for neutrons entering the
LEND SHEEN is comparable with that of the LP omni-directional
epithermal detector.

The best available map of the variation of the lunar HEE
flux, determined in this study and shown in Fig.~\ref{fig:hee},
differs from the input template for the lunar HEE component that was
estimated from LP results. This implies that the LEND SHEEN data
contain extra information that, in conjunction with Monte Carlo
neutron transport simulations, can be used to construct a global map
of the lunar hydrogen distribution (albeit at a
broad spatial resolution like LP). An ambiguity in the inferred
hydrogen abundance from LEE neutron measurements in non-polar regions
\citep{sly04,law06} will be absent from the HEE neutron case
\citep{john02}. Specifically, HEE neutrons are, unlike LEE neutrons,
unaffected by absorption \citep{law06}, but are still sensitive to
hydrogen variations \citep{gen03,law11a}.

A global map of the hydrogen abundance would provide a valuable, new
constraint on models for the lunar water cycle at low latitudes and a
useful consistency check of the high-altitude hydrogen maps from
LP. With the recent infra-red measurements indicating the existence of
surficial water at low latitudes and the LCROSS discovery of water in
a small patch of the Cabeus crater near to the lunar south pole, a
global hydrogen map would help to tie together these two separate
strands of evidence concerning lunar water. This study has shown that
data from the LEND experiment have the capacity to provide a global
hydrogen map, at a resolution similar to that which LP achieved near
to the poles. However, this conclusion relies upon a correct
interpretation of the observational data and understanding of the
instrumental capabilities. In order to exploit
the full potential of the LEND SHEEN data with confidence, one would need a
more detailed modelling of the detector response, with both position
and energy, than is currently available.

\acknowledgments

The Bartol Research Institute neutron monitor program is supported by
the United States National Science Foundation under grants ANT-0739620
and ANT-0838839, and by the University of Delaware Department of
Physics and Astronomy and Bartol Research Institute. V. Eke
acknowledges helpful discussions with C. Frenk and A. Jenkins.
D. Lawrence acknowledges the support of the NASA Lunar Science Institute.

\end{document}